\documentclass[a4paper,10pt]{article}

\usepackage{cite, amsfonts, amsthm,  fullpage}

\usepackage{amssymb}
\usepackage{amsbsy}
\usepackage{epsfig}
\usepackage{verbatim}

\newcommand{\Pf}{\mathop\mathrm{Pf}\nolimits}
\newcommand{\sgn}{\mathop\mathrm{sgn}\nolimits}
\newcommand{\Pa}{\mathop\mathrm{P}\nolimits}

\newcommand{\bt}{\mathbf{t}}

\theoremstyle{plain}

\newtheorem{Theorem}{Theorem}
\newtheorem{Lemma}{Lemma}
\newtheorem{Proposition}{Proposition}

\theoremstyle{remark}
\newtheorem{Remark}{Remark}

\def\l{\langle}
\def\r{\rangle}
\def\g{\Gamma}

\def\Tr{\mathrm {Tr}}
\def\tr{\mathrm {tr}}
\def\det{\mathrm {det}}

\def\diag{\mathrm {diag}}
\def\res{\mathop{\mathrm {res}}\limits_}

\def\bp{\begin{Proposition}\rm}
\def\ep{\end{Proposition}}
\def\bc{\begin{corollary}}
\def\ec{\end{corollary}}
\def\bl{\begin{Lemma}\em}
\def\el{\end{Lemma}}
\def\be{\begin{equation}}
\def\ee{\end{equation}}
\def\br{\begin{Remark}\rm\small}
\def\er{\end{Remark}}
\def\brs{\begin{remarks}.\\ \rm\
\begin{enumerate}}
\def\ers{\end{enumerate}\end{remarks}}
\def\bea{\begin{eqnarray}}
\def\eea{\end{eqnarray}}

\def\Tr{\mathrm {Tr}}
\def\tr{\mathrm {tr}}
\def\det{\mathrm {det}}

\def\sgn{\mathrm {sgn}}

\def\diag{\mathrm {diag}}

\def\res{\mathop{\mathrm {res}}\limits}

\def\&{&{\hskip -20pt}}

\newcount\YDcount\YDcount=0
\def\YDsize{10pt}

\def\YD#1{%
\ifnum#1=0
 \ifnum\YDcount=0 \ifx\varnothing\undefined\emptyset\else\varnothing\fi
 \else\vskip1.4pt\egroup\YDcount=0\fi
\else
 \ifnum\YDcount=0 \YDcount=1\vcenter\bgroup\vskip1pt
 \else\nointerlineskip\fi
 \vbox{\hrule\hbox{\vrule height\YDsize
 \loop\hskip\YDsize\vrule\ifnum\YDcount<#1\advance\YDcount1\repeat}\hrule
 \kern-0.4pt}\expandafter\YD
\fi}

\begin{document}
\author{ A. Yu.
Orlov\thanks{Institute of Oceanlogy, Nahimovskii Prospekt 36,
Moscow, Russia, email: orlovs55@mail.ru}}
\title{Deformed Ginibre ensembles and integrable systems}

\maketitle

\begin{abstract}

We consider three Ginibre ensembles (real, complex and quaternion-real) with a deformed measure and relate them
to the known integrable systems via presenting partition functions of these ensembles in form of fermionic expectation values.
We also introduce double deformed Dyson-Wigner ensembles and compare their fermionic representations with these of 
Ginibre ensembles.

\end{abstract}

\bigskip


\section{Introduction}
 
Ginibre ensembles play an important role in many statistical problems. They were introduced in 1965 by
Ginibre as non Hermitian analogues of famous Wigner-Dyson ensembles, namely, ensembles of random
real symmetric, Hermitian and quaternion self-dual random matrices, known also as orthogonal, unitary
and symplectic Wigner-Dyson ensembles. 
The non-Hermitian analogues are called real Ginibre, complex Ginibre and quaternion-real Ginibre ensemble respectively.

Let us recall that both Wigner-Dyson ensembles and Ginibre counterparts are Gauss ones. This was enough for  
the problems considered by physisists working in quantum chaos.
It is widely known that the deformation of the Gauss measure of each Wigner-Dyson ensemble makes them
tau-functions of integrable hierarchies where deformation parameters treated as the so-called higher times. 
First this link was established for the unitary ensemble which was identified with the 1D Toda lattice
tau function \cite{GMMMO}, and later also for the orthogonal and symplectic ones which were related to the so-called
Pfaff lattice \cite{AvM-Pfaff}. 

The origin for the deformation of WD ensembles were applications of these models to certain problems in physics 
(string theory and 2D quantum gravity and other application of summation over poly-angulated surfaces) and in mathematics
(few enumeration problems). At present there are no special motivation to consider deformations of the Ginibre ensembles,
however one may believe in future applications.

Wigner and Dyson had physical reasons to consider orthogonal, unitary and symplectic ensembles.
Physical systems are described by Hamiltonian operators which are Hermitian. Ginibre had no
physical reasons and introduced non-Hermitian analogues from "mathematical" point of view, having a hope that 
they may be of use in future. Indeed at present Ginibre ensembles are in a focus of attention in
many fields (such as
  quantum chromodynamics,
  dissipative quantum maps,
  scattering in chaotic quantum systems,
 growth processes,
 fractional quantum-Hall effect,
 Coulomb plasma,
 stability of complex biological and neural networks,
 directed quantum chaos in randomly pinned superconducting vortices,
 delayed time series in financial markets,
 random operations in quantum information theory, see a review article \cite{KhoruzhSommers} for details.)

 I am going to show that the partition function of the real and quaternion-real Ginibre ensembles are related to
 integrable systems similar to many other matrix models \cite{GMMMO}, \cite{KMMOZ}, \cite{Morozov}, \cite{MirMorSem}, 
 \cite{AvM-Pfaff}, \cite{L1} which may be referred as deformed Dyson-Wigner ensembles.
Here we use the so-called ``large'' BKP hierachy introduced in \cite{KvdLbispec} (named by authors ``charged'' or
``fermionic'' BKP) and the ``large'' 2-BKP hierarchy introduced in \cite{OST-I}. The last hierarchy is rather similar 
to 2-DKP (Pfaff-DKP) one introduces in \cite{T-09}\footnote{ Large BKP hierarchy includes the famous KP one as a particular 
reduction, while large 2-BKP includes the Toda lattice hierarchy \cite{UT}. In the present text we will refer these 
hierarchies as BKP and 2-BKP ones instead of the ``large BKP'' and ``large 2-BKP''. Make difference between these 
``large'' BKP hierarchy containing KP and the ``small'' (referred in \cite{KvdLbispec} as ``neutral'') one introduced in 
\cite{DJKM-BKP} and which is a subhierarchy of the KP hierarchy.}
We shall see that real and quaternion-real Ginibre ensemble both deformed in a appropriate way may be related
to these hierarchies where the higher times, namely, an integer $L$ and a pair of semi-infinite sets of numbers
$\bt=(t_1,t_2,\dots)$ and ${\bf s}=(s_1,s_2,\dots)$ plays the role of deformation parameters. 
We say that this is ``double'' deformed ensembles because two sets of parameters is used. 

To be more precise it is only the case where $L\ge 0$ and ${\bf s}=0$ may be related to what is called Ginibre ensembles.
To consider more general deformations 
we need to introduce {\em restricted} (real, quaternion-real, complex) Ginibre ensembles where the space of random 
matrices consists of only invertable (real, quaternion-real, complex)  matrices.
Then we prove that partition functions of deformed (real and quaternion-real) Ginibre ensembles are BKP tau functions, where $\bt$
and $L\ge 0$ are deformation parameters and BKP higher times. While partition functions of double deformed restricted 
(real and quaternion-real) ensembles are 2-BKP tau functions, where now the set $L,\bt,{\bf s}$ are deformation parameters and 
higher times.

A fermionic representation for the tau functions will be written down and compared with the representation of appropriately
deformed classical Dyson-Wigner ensembles, where an additional set of parameters ${\bf s}$ is added.
The complex Ginibre ensemble will be deformed with the help of four sets of parameters, $\bt,\bt',{\bf s},{\bf s}'$, and 
related to the two-component Toda lattice. 

\section{Gauss Real Ginibre ensemble and its deformation}

{\bf Real Hermitian ensemble}. First let us remind that the ensemble of real Hermitian matrices (also known as 
Dyson-Wigner orthogonal ensemble) is given by the following partition function
\be\label{d-OE}
I_N^{d-OE}({\bf t})=\int d\mu(X,{\bf t})\,,\quad 
d\mu(X,{\bf t})=  e^{-\frac 12 \tr (X^2)+\tr V(X,{\bf t})}\prod_{i\ge j}\,dX_{ij}
\ee
  \be\label{V}
V(x,\bt):={\sum_{n=1}^\infty  x^n t_n}
 \ee
where $X$ is $N$ by $N$ real Hermitian matrix and and ${\bf t}=(t_1,t_2,\dots)$ is a set of parameters 
which describe the deviation of the probability measure from the Gauss one. 

Let us restrict the space of our symmetric matrices to invertable symmetric matrices. Then more general deformation 
of the measure may be considered as follows
\be\label{dd-OE}
I_N^{dd-OE}(L,{\bf t},{\bf s})=\int d\mu(X,L,{\bf t},{\bf s}) \,,\qquad 
d\mu(X,L,{\bf t},{\bf s})=\det X^L e^{-\tr V(X^{-1},{\bf s})}\, d\mu(X,\bt)
\ee
where ${\bf s}$ and an integer $L$ is the collection of new deformation parameters. Dyson-Wigner Gauss ensemble will be referred
as G-OE, ensembles (\ref{d-OE}) and (\ref{dd-OE}) respectively as d-OE and dd-OE.

It was found by M.Adler and P. van Moerbeke in \cite{AvM-Pfaff} that the ensemble (\ref{d-OE}) may be related to the so-called
Pfaff Toda lattice \cite{AMS}. Later in \cite{L1} J. van de Leur found that the orthogonal ensemble (\ref{d-OE}) may be 
also identified with a tau function of the ``large'' BKP tau function introduced in \cite{KvdLbispec}. Below we shall 
identify the ensemble (\ref{dd-OE}) with a ``large'' 2-BKP tau function introduced in \cite{OST-I}.

In terms of eigenvalues $x_1,\dots,x_N $ of the Hermitian matrix $X$ the integral (\ref{dd-OE}) may be written as
\be\label{dd-OEeigen}
I_N^{dd-OE}(L,{\bf t},{\bf s})=a_N\int_{x_1>\cdots>x_N} \Delta(x_1,\dots,x_N)
\prod_{i=1}^N x_i^L
 e^{-\frac 12 x_i^2+ V(x_i,{\bf t})- V(x_i^{-1},{\bf s})} d x_i
 \ee
with some constant $a_N$ related to the volume of the orthogonal group $O(N)$. Deformation parameters ${\bf t}$ and 
${\bf s}$ are considered to be chosen in a way that the integral (\ref{dd-OEeigen}) is convergent.

{\bf Real non-Hermitian ensemble}. Let us turn to the so-called Ginibre ensembles. 
Gauss real Ginibre ensemble, also known as Gauss Ginibre orhtogonal ensemble (G-GinOE) is defined on the space of 
real matrices by assigning Gauss probability 
measure to each entry with the same variance:
\be\label{G-GinOE}
I_N^{G-GinOE}=\int d\mu(X)\,,\quad 
d\mu(X)=\prod_{i,j}\,e^{-\frac 12 X_{ij}^2}dX_{ij}=e^{-\frac12 \tr (XX^\dag)}\prod_{i,j}\,dX_{ij}
\ee
where $X$ is $N$ by $N$ matrix with real entries.
Measure (\ref{G-GinOE}) is invariant under the orthogonal transformations of matrices $X$.

The so-called elliptic deformation of G-GinOE measure as 
\be\label{G-GinOE-ell}
d\mu(X,a)=e^{-a\tr X^2}d\mu(X)
\ee
where $a$ is the deformation parameter was studied in  \cite{KhoruzhSommers} \footnote{ It is also called
the elliptic deformation of G-GinOE because in the large $N$ limit the parameter $a$ describes a deformation of
a circle equilibrium domain of the eigenvalues in complex plane to an elliptic form. } .

We shall consider the following deformation of the G-GinOE ensemble which we will refer as d-GinOE and which includes 
the elliptic deformation as a particular case
 \be\label{d-GinOE}
I_N^{d-GinOE}(L,{\bf t})=\int \,d\mu(X,L,{\bf t})\,,\qquad
d\mu(X,L,{\bf t})=\det X^L \, e^{\tr V(X,\bt)}\, d\mu(X)
 \ee
where $V$ is given by (\ref{V}) and where $L=0,1,2\dots$ and sets of numbers $\bt=(t_1,t_2,\dots)$ are deformation parameters.
We assume that $\bt$ are chosen in a way that the partition function $I_N^{d-GinOE}(L,{\bf t})$ is finite.
 \br \label{regul}
 However even in the case where the integral (\ref{d-GinOE}) is divergent itself, logarithmic derivatives with respect to
the deformation parameters may be finite. How it works in the models of normal matrices, see in \cite{regularization}. 
 \er
 
 Now let us consider the restricted Ginibre ensemble consisting of real invertable matrices and deform its measure 
 as follows
 \be\label{dd-GinOE}
I_N^{dd-GinOE}(L,{\bf t},{\bf s})=\int \,d\mu(X,L,{\bf t},{\bf s})\,,\qquad
d\mu(X,L,{\bf t},{\bf s})=\det X^L \, e^{\tr V(X,\bt)-\tr V(X^{-1},{\bf s})}\, d\mu(X)
 \ee
where $X$ are real invertable $N$ by $N$ matrices and where $V$ is given by (\ref{V}) and
where an integer $L$ and sets of numbers $\bt=(t_1,t_2,\dots)$, ${\bf s}=(s_1,s_2,\dots)$ are deformation parameters.
This ensemble will be called double deformed restricted Ginibre orthogonal ensemble (dd-GinOE). Again we suppose that deformation 
parameters are chosen in a way which provides the existence of the integral $I_N^{dd-GinOE}(L,{\bf t},{\bf s})$.
Consider the following example. If one takes $\bt \to \bt- \sum_{i=1}^{N_1}{a_i} [p_i]$ and
${\bf s} \to {\bf s}- \sum_{i=1}^{N_1}{b_i} [q_i]$, namely
 \be\label{Miwa-choice}
t_n\to t_n -\frac 1n \sum_{i=1}^{N_1}{a_i} p_i^{n}  ,\qquad s_n\to s_n-\frac 1n \sum_j^{N_2} {b_j} q_j^{
n}
 \ee
then the integral (\ref{dd-GinOE}) reads as
 \be\label{det-averages}
\int \,\prod_{i}^{N_1}\det\left(1-p_iX \right)^{a_i}
\prod_j^{N_2} \det\left(1-q_iX^{-1} \right)^{-b_i}\, d\mu(X,L,{\bf t},{\bf s})
 \ee

Let prove that the integrals (\ref{d-GinOE}) and (\ref{dd-GinOE}) may be identified with tau functions of the 
"large" BKP hierarchy and 2-BKP respectively. 
Then the size of matrices $N$ and the set $L,\bt,{\bf s}$ play the role of 
the so-called higher times of this integrable hierarchy. 

To do it we should re-write integral (\ref{dd-GinOE}) as the integral over eigenvalues. Let us remind that for the Gauss 
real Ginibre ensemble (namely, for the case $\bt={\bf s}=0$, $L=0$) this problem was solved
in \cite{Sommers08}, see also a review article \cite{KhoruzhSommers}. In very short it was done as follows.
After applying the Schur decomposition $X=U(\Lambda + \Delta)U^\dag$,
where $U$ is orthogonal, $\Lambda$ is block-diagonal and $\Delta$ has nonzero blocks only above $\Lambda$ 
they arrive at
$$
d\mu(X) =  e^{-2\Tr(\Delta\Delta^\dag + \Lambda\Lambda^\dag)}  |D\Delta| |D\Lambda |
|{\prod}{'}\left(U^{-1}dU \right)_{ij}{\left( \lambda_i-\lambda_j\right)\over {2\pi}}|
$$
with the dashed product running over non-zero entries in the lower triangle of
$U^{\dag} dU$. The set of the eigenvalues $\lambda_i$, $i=1,\dots,N$ of a real matrix $X$ consists of real numbers and of 
complex conjugated
pairs, denoted below as $x_i$ and $(z_i,{\bar z}_i)$ respectively. 
Thus, the matrix $\Lambda=\diag(\lambda_1,\dots,\lambda_N)$ may be written as 
$\diag(z_1,{\bar z}_1,\dots,z_k,{\bar z}_k, x_1,\dots, x_{N-2k})$ with certain $k$. After integration over matrices $\Delta$ 
and $U$ they come to the integral over eigenvalues only
which includes integral over real values $x_i$ and over complex eigenvalues $z_j,{\bar z}_j$. Finely, after certain computations 
\cite{Sommers08} the Gauss real Ginibre ensemble may be written as
 \be\label{I^{G-GinOE}_{N}}
I_N^{G-GinOE}=b_N\cdot
\sum_{k=0}^{\left[\frac N2 \right]}\int_{\mathbb{M}_{2k,N-2k}} \,
\Delta_{2n}(z_1,{\bar z}_1,\dots ,z_k,{\bar z}_k,x_{1},\dots,x_{N-2k})\,
d\Omega^C_{2k}\,d\Omega^R_{N-2k}
 \ee
where the integration domain $\mathbb{M}_{2m,N-2m}$ is as follows  $\Re z_1>\cdots >\Re z_m$, $x_{2m+1}>\cdots>x_N$, 
$z_i\in\mathbb{C}_+$ (upper half-plane), $x_i\in\mathbb{R}$, and
where
 \be
  d\Omega^C_{2m}({\bf z},{\bar{\bf z}})=
\prod_{i=1}^m \mbox{erfc}\left(\frac{|z_i-{\bar z}_i|}{\sqrt 2} \right) 
 e^{-\Re z_i^2} d^2 z_i\,,\qquad
 d\Omega^R_{N-2m}({\bf x})=
\prod_{i=2m+1}^N 
 e^{-\frac 12 x_i^2} d x_i
 \ee
 with $\mbox{erfc}(z)=\frac{2}{\sqrt{\pi}}\int^{\infty}_z e^{-x^2}$.
Here the factor $c_N$ absorbs integrals over $\Delta$ and over $U$.
Let us note that the Gauss case is a special case of (\ref{dd-GinOE}): $I_N^{G-GinOE}=I_N^{dd-GinOE}(0,0,0)$.
The factor $b_N$ is independent of $L,\bt,{\bf s}$ and was evaluated for GinOE in \cite{Sommers08}.

Now we notice that the deformation (\ref{dd-GinOE}) results in the multiplication of the measure $d\mu(X)$ by a factor which
depends only on eigenvalues of the matrix $X$
 \be
d\mu(X)  \to d\mu(X,\bt,L,{\bf s})=d\mu(X) \prod_{i=1}^N \lambda_i^Le^{V(\lambda_i,\bt)-V(\lambda_i^{-1},{\bf s})}
 \ee
where $\lambda_i$ are eigenvalues of $X$ and $V$ is given by (\ref{V}). Then we introduce
 \be
  d\Omega^C_{2m}\to d\Omega^C_{2m}(\bt,L,{\bf s})=
\prod_{i=1}^m \mbox{erfc}\left(\frac{|z_i-{\bar z}_i|}{\sqrt 2} \right) 
|z_i|^{2L} e^{2\Re V(z_i,\bt)-2\Re V(z_i^{-1},{\bf s})-\Re z_i^2} d^2 z_i
 \ee
 \be
 d\Omega^R_{N-2m}\to d\Omega^R_{N-2m}(\bt,L,{\bf s})=
\prod_{i=2m+1}^N x_i^L e^{V(x_i,\bt)-V(x_i^{-1},{\bf s})-\frac 12 x_i^2} d x_i
 \ee

Let us introduce
 \be\label{J_N-r}
J_N(L,{\bf t},{\bf s};\alpha)=
\sum_{k=0}^{\left[\frac N2 \right]}\alpha^k \int_{\mathbb{M}_{2k,N-2k}} \,
|z_i|^{2L}
\Delta_{2n}(z_1,{\bar z}_1,\dots ,z_k,{\bar z}_k,x_{1},\dots,x_{N-2k})\,
d\Omega^C_{2k}(\bt,L,{\bf s})\,d\Omega^R_{N-2k}(\bt,L,{\bf s})
 \ee
 We note that for $\alpha\to 0$ (up to a constant independent of the variables $L,{\bf t},{\bf s}$) this expression is 
 equal to the partition function of the deformed Dyson-Wigner orthogonal ensemble (\ref{d-GinOE}). When $\alpha=1$ this is 
 the partition function for the deformed real Ginibre ensemble (\ref{d-GinSE}). 
 \be \label{JI-d}
J_N(L,{\bf t},{\bf s}=0;\alpha=1) =b_N I_N^{d-GinOE}(L,{\bf t})
 \ee
 \be \label{JI-dd}
J_N(L,{\bf t},{\bf s};\alpha=0)=a_N I_N^{dd-OE}(L,{\bf t},{\bf s}) \,,\qquad 
J_N(L,{\bf t},{\bf s};\alpha=1) =b_N I_N^{dd-GinOE}(L,{\bf t},{\bf s})
 \ee
Independent of deformation parameters factors $a_N$ and $b_N$ may be found respectively in \cite{Mehta} and \cite{KhoruzhSommers}.

{\bf Tau functions and matrix ensembles}. Let us remind the fermionic construction of the so-called tau functions \cite{JM}.
Namely here we present a general fermionic expression for the 2-BKP tau function and relate it to ensembles 
(\ref{dd-OE}) and (\ref{dd-GinOE}). In the next section it will be also related to a deformed symplectic Dyson-Wigner ensemble 
and to a deformed quaternion-real Ginibre ensemble.

 The very notion of tau function and it's fermionic construction was introduces by Kyoto school. Here we use it in a 
 version suggested in \cite{KvdLbispec} where an additional fermioc mode $\phi$ was added. see Appendix \ref{Fermions}.
 
Following \cite{JM} we consider
 \be
\g_+(\bt)=\exp \sum_{n=1}^\infty t_n\sum_{i\in\mathbb{Z}} \psi_i\psi^\dag_{i+n},\qquad
\g_-({\bf s})=\exp \sum_{n=1}^\infty s_n\sum_{i\in\mathbb{Z}} \psi_i\psi^\dag_{i-n}
 \ee

We shall need the following equality 
 \be\label{gamma-psi-gamma}
\g_+(\bt) \psi(z)\g_-({\bf s})=c(\bt,{\bf s})e^{V(z,\bt)-V(z^{-1},{\bf s}))}\g_-({\bf s}) \psi(z) \g_+(\bt)
 \ee
where
 \be\label{c}
c(\bt,{\bf s})=\exp \sum_{n=1}^\infty nt_ns_n
 \ee

 The 2-BKP tau function may be presented in form of the following fermionic expectation value
  \be\label{TauN} 
\tau_N(L,{\bf t},{\bf s})= \l N+L|\,\g_+({\bf t})\, e^{\Phi} 
\,{\g}_-({\bf s})\,|L\r
 \ee
 where $\Phi$ is any quadratic expression in fermionic modes $\psi_i,\psi^\dag_i,\phi$, see Appendix \ref{Fermions}. Variables $N,L,\bt,{\bf s}$
 plays the role of the 2-BKP higher times. Tau function (\ref{TauN}) solves 2-BKP Hirota equations, see \cite{OST-I}
 and Appendix \ref{Hirota}.
 
 \br\label{2BKP-BKP}
 The 2-BKP tau function is also a certain BKP tau function \cite{KvdLbispec} with 
 respect to the variables $N,L,\bt$ and a certain BKP tau function with respect to the variables $N,L,{\bf s}$ (this explains 
 the name 2-BKP). 
 
 Tau function (\ref{TauN}) contains as a particular cases: tau functions of the Toda lattice 
 hierarchy \cite{JM}, \cite{UT}, the charged BKP tau 
 function \cite{KvdLbispec} and 2-DKP (``Pfaff DKP'') tau function \cite{T-09}. Each of these cases may be obtained by 
 specifying the tau function (\ref{TauN}), namely, by specifying $\Phi$ and ${\bf s}$, see Appendix \ref{reductions}.
 \er

\begin{Theorem}
We have
 \be\label{Theorem1} 
(-)^{NL}c(\bt,{\bf s}) J_N(L,{\bf t},{\bf s};\alpha)= \l N+L|\,\g_+({\bf t})\, e^{\alpha \Phi_c}e^{\Phi_r} 
\,{\g}_-({\bf s})\,|L\r
 \ee
where
 \be
\Phi_c=\int_{\mathbb{C}_+} \mbox{erfc}\left(\frac{|z-{\bar z}|}{\sqrt 2} \right) 
\psi(z)\psi({\bar z})e^{-\Re z^2}d^2z 
 \ee
is the integral over the upper half of complex plane and where
 \be\label{Pr}
\Phi_r=\frac 12 \int_{\mathbb{R}}\int_{\mathbb{R}} 
\sgn(x_1-x_2) \psi(x_1)\psi(x_2)e^{-\frac 12 x_1^2-\frac 12 x_2^2}dx_1dx_2
+{\sqrt 2}\int_{\mathbb{R}} \psi(x)\phi e^{-\frac12 x^2} dx
 \ee
Therefore thanks to (\ref{JI-d}) and Remark \ref{2BKP-BKP} $(-)^{NL}I_N^{d-GinOE}(L,{\bf t})$ is a 
tau function of the ``large'' BKP  
while thanks to (\ref{JI-dd}) both $(-)^{NL}c(\bt,{\bf s})I_N^{dd-OE}(L,{\bf t},{\bf s})$ and 
$(-)^{NL}c(\bt,{\bf s})I_N^{dd-GinOE}(L,{\bf t},{\bf s})$ of 
the ``large'' 2-BKP ones.

\end{Theorem}

The sketch of the proof of (\ref{Theorem1}). We transform the vacuum expectation value in the right hand side to get the 
integral in the left hand side as follows. 
First we use  Taylor expansion of the exponentials $e^{ \alpha \Phi_c}$ and 
$e^{\beta \Phi_r}$ 
where only terms of the order $\alpha^{N-2k}\beta^k$, $k=0,\dots,\left[ \frac N2 \right]$ are non-vanishing between 
$\l N+L|$ and $|L\r$. The term $e^{\alpha \Phi_r}$ should be
 considered in a way similar to \cite{L1} (see Appendix). Notice that the Taylor expansion of the exponentials of integrals
yields the integration domain $ \mathbb{M}_{2k,N-2k}$ when we re-write product of pairwise integrals as a $N$-fold integral. 
Next we send $\g$ to the right and $\g^*$ to the left
taking into account relations
$\l N+L|{\g}^*({\bf s})=\l N+L|$ and $\g(\bt)|L\r=|L\r$ and using (\ref{gamma-psi-gamma}). In this way we get rid of 
operators $\g$ and $\g^*$ and instead obtain factors
$e^V$ inside integrals, the integrals are still between $\l N+L|$ and $|L\r$. At last we get rid of fermions
thanks to
 \be
\l N+L|\psi(z_1)\cdots \psi(z_N)|L\r=z_1^L\cdots z_N^L \Delta_N({\bf z}),\quad 
\Delta_N({\bf z}):=\prod_{i<j\le N}(z_i-z_j)
 \ee
obtained by a simple direct calculation. We obtain (\ref{Theorem1}).

Now the fermionic expectation value $\l N+L|\,\g({\bf t})\, e^{\alpha \Phi_c}e^{\Phi_r} 
\,{\g}^*({\bf s})\,|L\r$ is an example of tau function with respect to the variables $N,L,\bt$ introduced in \cite{KvdLbispec} 
where Hirota equations for such tau functions were written down. By symmetry it is also tau function with respect to the 
variables $N,L,{\bf s}$. Such ("coupled" BKP, or, the same 2-BKP) tau functions were considered in \cite{OST-I}.

Thus we establish that the partition function of the deformed quaternion-real Ginibre ensemble (\ref{d-GinOE}) is the subject to
the theory of integrable systems.

\section{Quaternion-real Ginibre ensemble and its deformation}

Each $N \times N$ quaternion matrix (or, the same, a matrix with quaternion entries) may be viewed as $2N\times 2N$ matrix if 
the quaternions
$e_n$, $n=1,2,3,4$, are realized as $2\times 2$ matrices:  
\be\label{e}
e_0= \pmatrix{1 & 0 \cr 0 & 1}\,,\quad e_1= \pmatrix{i & 0 \cr 0 & -i}\,,\quad
e_2= \pmatrix{0 & 1 \cr -1 & 0}\,,\quad e_3= \pmatrix{0 & i \cr i & 0}
\ee
Then quaternion-real matrix is a quaternion matrix where each entry is a linear combination
of quaternions with real coefficients. Being viewed as $2N\times 2N$ matrix a quaternion-real matrix $X$ may be
recognized via the property 
 \be\label{quat2N}
EXE=-{\bar X}
 \ee
 where the bar means complex conjugate and where $E$ is the block-diagonal $2N\times 2N$ matrix with matrices 
$\pmatrix{0 & 1 \cr -1 & 0}$ on the main diagonal; 
this follows from the explicit expression (\ref{e}) and from $e_2e_je_2=e_j$, $j=1,3$ and from $e_2e_je_2=-e_j$, $j=0,2$.
Below we shall treat quaternion-real matrices as $2N\times 2N$ matrices with the defining property (\ref{quat2N}).

{\bf Quaternion-real Hermitian ensemble}. First we recall the case where in the quaternion-real matrix $X$ is Hermitian 
(in this case $X$ is called self-dual). The ensemble of random self-dual matrices is called symplectic ensemble, the Gauss 
symplectic ensemble (G-SE) is called Dyson-Wigner symplectic ensemble. We will consider a deformed  symplectic 
ensemble (d-SE) as follows
 \be
I_N^{d-SE}(L,{\bf t})=\int d\mu(X,L,{\bf t}) \,,\qquad 
d\mu(X,L,{\bf t})= e^{-x^2+2\tr V(X,\bt)}\, d\mu(X)
 \ee
where $d\mu(X)$ is the measure on the space of self-dual matrices (for details see \cite{Mehta}).
As in the orthogonal ensemble case we may consider the subspace of invertable matrices and consider double
deformed restricted ensemble of quaternion-real matrices (dd-SE)
 \be\label{dd-SE}
I_N^{dd-SE}(L,{\bf t},{\bf s})=\int d\mu(X,L,{\bf t},{\bf s}) \,,\qquad 
d\mu(X,L,{\bf t},{\bf s})=\det X^L e^{-x^2+2\tr V(X,\bt)-2\tr V(X^{-1},{\bf s})}\, d\mu(X)
 \ee
where we add the deformation parameters ${\bf s}=(s_2,s_2,\dots$ and integer $L$.
Being re-written as an integral over eigenvalues of Hermitian quaternion real matrices $X$ \cite{Mehta} it is as follows
 \be\label{dd-SEeigen}
I_N^{dd-SE}(L,\bt,{\bf s}) =d_N\int \left(\Delta(x_1,x_2,\dots,x_N)\right)^4 \prod_{i=1}^N x_i^{2L}
 e^{-x_i^2+ 2V(x_i,{\bf t})- 2V(x_i^{-1},{\bf s})} d x_i
 \ee
 where $x_1,x_1,x_2,x_2\dots,x_N,x_N$ are eigenvalues of the self-dual random $2N $ by $2N$ matrix $X$ and
 the factor $d_N$ does not depend on deformation patameters.

{\bf Quaternion-real non-Hermitian ensemble}. Then Gauss quaternion-real Ginibre ensemble, also known as (Gauss) Ginibre
symplectic ensemble (G-GinSE) is defined on the space of 
quaternion-real matrices by assigning the same Gauss probability measure to each entry:
\be\label{G-GinSE}
I_N^{G-GinSE}=\int d\mu(X)\,, \qquad d\mu(X)=e^{-\frac 12 \tr (XX^\dag)}\prod_{i,j}\,dX_{ij}
\ee
where $X$ is treated as $2N$ by $2N$ matrix.
Measure (\ref{G-GinSE}) is invariant under the symplectic transformations of matrices $X$.

\br
 A Hermitian quaternion-real matrix is called quaternion self-dual matrix. Ensemble (\ref{G-GinSE}) is non-Hermitian analogue of 
the ensemble of random quaternion self-dual matrices (known as Dyson-Wigner symplectic ensemble), see \cite{Mehta} for details.
\er

The elliptic deformation of G-GinSE measure \cite{KhoruzhSommers} is quite similar to (\ref{G-GinOE-ell}):
$
d\mu(X) \to e^{-a\tr (X^2+(X^\dag)^2}d\mu(X)
$
where $a$ is a deformation parameter.

The partition function for the deformed quaternionic-real Ginibre ensemble (d-GinSE) will be defined as
 \be\label{d-GinSE}
I_N^{d-GinSE}(L,{\bf t})=\int \, d\mu(X,L,{\bf t}),\quad
d\mu(X,L,{\bf t}):=\det X^L \, e^{\tr V(X,\bt)}\, d\mu(X)
 \ee
where a non-negative integer $L$ and the set of numbers ${\bf t}=(t_1,t_2,\dots)$ are deformation parameters.

On the space of invertable quaternion-real matrices we consider double deformed Ginibre ensemble
 \be\label{dd-GinSE}
I_N^{dd-GinSE}(L,{\bf t},{\bf s})=\int \, d\mu(X,L,{\bf t},{\bf s})\,,
\quad d\mu(X,L,{\bf t},{\bf s})=\det X^L \, e^{\tr V(X,\bt)-\tr V(X^{-1},{\bf s})}\, d\mu(X)
 \ee
 where an integer $L$ and and the sets  ${\bf t}=(t_1,t_2,\dots)$, ${\bf s}=(s_1,s_2,\dots)$ are deformation parameters.
 Here we assume that the integral is either convergent or regularized, see Remark \ref{regul}.
 
Notice that if one takes
 \be\label{Miwa-choice-sympl}
t_n\to t_n-\frac 1{2n} \sum_{i=1}{a_i} p_i^{n}  ,\qquad s_n\to s_n-\frac 1{2n} \sum_j {b_j} q_j^n
 \ee
then the integral (\ref{d-GinSE}) reads as
 \be
\int \,\det X^L \,\prod_{i}\det\left(1-p_iX \right)^{a_i}\prod_j \det\left(1-q_iX^{-1} \right)^{-b_i}\, 
d\mu(X,L,{\bf t},{\bf s})
 \ee

As in the case of real matrices all complex eigenvalues occur in complex conjugated pairs. Real eigenvalues have multiplicities 
no less than two. A calculation similar to calculation for real matrices yields \cite{KhoruzhSommers} (see also 
\cite{Kan2005}, \cite{For2007}, \cite{Mehta})
 \be\label{I_N-d-GinSE-eigen}
I_N^{dd-GinSE}(L,{\bf t},{\bf s})=
 \ee
 \[
c_N\int_{\mathbb{M}_n} \,\Delta_{2n}(z_1,{\bar z}_1,z_2,{\bar z}_2,\dots,z_n,{\bar z}_n)
\prod_{i=1}^n|z_i-{\bar z}_i||z_i|^{2L}e^{V(z_i,\bt)+V({\bar z}_i,\bt)-V(z_i^{-1},{\bf s})-V({\bar z}_i^{-1},{\bf s})- |z_i|^2} d^2 z_i
 \]
with some constant $c_N$ where the integration domain $\mathbb{M}_n$ consists of the sets of ${\bf z}$ where 
$\Re z_i>\Re z_{i+1}$ and $\Im z_i > 0$,  and $V$ is given by (\ref{V}).

\begin{Theorem}
Introduce
 \be\label{Theorem2} 
c(\bt,{\bf s}) J_N(L,{\bf t},{\bf s},\alpha,\beta)= \l 2N+L|\,\g_+({\bf t})\, e^{ \alpha\Phi_c} e^{ \beta\Phi_r} 
\,{\g}_-({\bf s})\,|N\r
 \ee
where
 \be\label{P}
\Phi_c^q =  \int_{\mathbb{C}_+} (z-{\bar z})
\psi(z)\psi({\bar z})e^{- |z|^2}d^2z 
 \ee
  \be\label{Phi-q-r}
\Phi_r^q = \int_{\mathbb{R}} 
\frac{\partial \psi(x)}{\partial x}\psi( x)e^{-x^2}dx 
 \ee 
is the integral over the upper half of complex plane. Then
 \be\label{d-GinSE-IJ}
 I_N^{d-GinSE}(L,{\bf t})= d_N J_N(L,{\bf t},{\bf s}=0,\alpha=1,\beta=0)
 \ee
 \be\label{dd-GinSE-IJ}
 I_N^{dd-GinSE}(L,{\bf t},{\bf s})= d_N J_N(L,{\bf t},{\bf s},\alpha=1,\beta=0)\,,\quad
 I^{dd-SE}(L,\bt,{\bf s})= c_N J_N(L,{\bf t},{\bf s},\alpha=0,\beta=1)
 \ee
Therefore $I_N^{d-GinSE}(L,{\bf t})$ is an example of the large ("fermionic") BKP tau function of \cite{KvdLbispec} with 
respect to the variables $N,L,\bt$. Then $c(\bt,{\bf s})I_N^{dd-GinSE}(L,{\bf t},{\bf s})$ and 
$c(\bt,{\bf s})I_N^{dd-SE}(L,{\bf t},{\bf s})$ are examples of the 
large 2-BKP tau function considered in \cite{OST-I} with respect
to the variables $N,L,\bt,{\bf s}$ and of large 2-DKP tau function introduced in \cite{T-09}, 
see Appendix \ref{reductions}.
\end{Theorem}
Let us note that the presentation with ${\bf s}=0$, namely, the representation for the symplectic Dyson-Wigner ensemble 
(where $\alpha=0$) was found by J. van de Leur in \cite{L1}.

\subsection{On perturbation series in deformation parameters}

In \cite{OST-I}   we have considered the following series over partitions
 \be
 \label{S}
\sum_{\lambda\in\Pa \atop \ell(\lambda)\le N}\,{\bar A}_{h(\lambda)}(L)
\,s_\lambda(\bt)
 \ee 
where $s_\lambda$ is the Schur function (see Appendix \ref{Schur}) and ${\bar A}_{h(\lambda)}(L)$ is a certain Pfaffian, see
Appendix \ref{Pfaffians}. 
Such series one obtains for many kinds of matrix integrals, see \cite{OST-II}, for instance for integrals over
symplectic and orthogonal groups, see Appendix \ref{integrals-over}. The notations are as follows.
  $\lambda=(\lambda_1,\dots,\lambda_N)$, $\lambda_1\ge \dots \lambda_N\ge 0$ is a partition, see \cite{Mac} and
$h(\lambda)=\lambda_i-i+N$ are the so-called shifted parts of $\lambda$. The factors ${\bar A}_h$  on the  right-hand 
side of (\ref{S}) are   determined in terms a pair $(A, a)=:{\bar A}$ where $A$ is an infinite skew symmetric matrix and $a$
an infinite vector.  For a set $h=
(h_1,\dots,h_N )$, $h_1>\cdots > h_N$ the numbers
${\bar A}_h$ are defined as the Pfaffian of an antisymmetric $2n
\times 2n$ matrix ${\tilde A}$  as follows:
  \be
  \label{A-c}
{\bar A}_{h}(L):=\,\Pf[{\tilde A}]
  \ee
where for $N=2n$ even
  \be
  \label{A-alpha-even-n}
{\tilde A}_{ij}=-{\tilde A}_{ji}:=A_{h_i+L,h_j+L},\quad 1\le
i<j \le 2n
  \ee
and for $N=2n-1$ odd
 \be \label{A-alpha-odd-n} {\tilde
A}_{ij}=-{\tilde A}_{ji}:=
 \cases{
A_{h_i+L,h_j+L} \qquad {\rm if }\quad 1\le i<j \le 2n-1 \cr
a_{h_i+L} \qquad \quad \, \,  {\rm if } \quad 1\le i < j=2n }  .
  \ee  
In addition we set ${\bar A}_0 =1$.

Having the fermionic representation it is straightforward to write down the perturbation series of partition functions
of Ginibre ensembles in the Schur functions like it was done in other cases 
(for instance see \cite{MirMorSem}, \cite{OS}, \cite{HO2006}). 

(I) First let us do it for the quaternion 
real Ginibre ensemble. Re-writing
 \be
 \int_{\mathbb{C}_+} (z-{\bar z})
\psi(z)\psi({\bar z})e^{-|z|^2}e^{-V(z^{-1},{\bf s})-V({\bar z}^{-1},{\bf s})}d^2z = 
\sum_{n,m} A_{nm}\psi_n\psi_m
 \ee
 where $A_{nm}$ is a moment matrix
  \be\label{quat-r-moment}
 A_{nm}^{GinSE}({\bf s})=\int_{\mathbb{C}_+} z^n{\bar z}^m (z-{\bar z})e^{-|z|^2-V(z^{-1},{\bf s})-V({\bar z}^{-1},{\bf s})}d^2z ,
 \qquad a_n^{GinSE}=0,\quad n,m=1,\dots,N 
  \ee
  we reduce problem to the solved in \cite{OST-I}. We obtain the following series in the Schur functions (see \cite{Mac} 
  for definitions and details) $s_\lambda$ 
  (please, do not mix with the deformation parameters $s_i$):
  \be
  I_N^{dd-GinSE}(L,{\bf t},{\bf s})=\sum_{\lambda\atop \ell(\lambda)\le 2N} {\bar A}^{GinSE}_{\{h\}}(L,{\bf s})s_\lambda(\bt)
  \ee

  In case ${\bf s}=0$ we get $A_{m,n}=-A_{n,m}=n!\delta_{m+1,n}$.

(II)  For DW  symplectic ensemble the formula is the same, 
  but now instead of (\ref{quat-r-moment}) we have 
   \be\label{symplDW-moment}
 A_{nm}^{SE}({\bf s})=\frac{n-m}{2}\int_{\mathbb{R}} x^{n+m-1} e^{-x^2-2V(x^{-1},{\bf s})}dx ,
 \qquad a_n^{SE}=0,\quad n,m=1,\dots,N 
  \ee

 (III)  For the DW orthogonal ensemble we have in case $N$ is even the formula is the same, However for $N$ odd we need to define
   $A_{\{h\}}$.
  \be\label{realDW-moment}
 A_{nm}^{OE}({\bf s})=\int_{\mathbb{R}}\int_{\mathbb{R}} 
 x^{n}y^m \sgn(x-y)e^{-\frac 12 x^2-\frac 12 y^2-V(x^{-1},{\bf s})-V(y^{-1},{\bf s})}dxdy ,
  \ee
  \be
  a_n^{OE}({\bf s})=\int_{\mathbb{R}} x^n e^{-V(x^{-1},{\bf s})}dx
  \ee

 (IV) For the real Ginibre ensemble we have
 
 \be\label{r-moment}
   A_{nm}^{GinOE}({\bf s})= \int_{\mathbb{C}_+} z^n{\bar z}^m \mbox{erfc}\left(\frac{|z-{\bar z}|}{\sqrt 2} \right)  
 e^{-\Re z^2 -V(z^{-1},{\bf s})-V({\bar z}^{-1},{\bf s})}d^2z
   \ee
  \[
 + \int_{\mathbb{R}}
 \int_{\mathbb{R}} x^{n}y^m \sgn(x-y)e^{-\frac 12 x^2-\frac 12 y^2-V(x^{-1},{\bf s})-V(y^{-1},{\bf s})}dxdy,
  \]
  \be
  a_n^{GinOE}({\bf s})=\int_{\mathbb{R}} x^n e^{-V(x^{-1},{\bf s})}dx
  \ee
  For $N$ even (which occurs in the cases of DW orthogonal and in real Ginibre ensembles) the formula for ${\bar A}_{\{h\}}$ is 
  slightly more spacious, see \cite{OST-I}. 
  
  Let us also note that in all cases where ${\bf s}=0$ (namely, for d-OE, d-GinOE, d-SE, d-GinSE ensembles) 
  moments can be explicitly evaluated in a straightforward way.

\section{On Pfaffian formulae}

Various Pfaffian formulae are known in the study of non-Hermitian matrices, see \cite{Kanzieper}, \cite{KhoruzhSommers}.
These formulae, and perhaps some new ones, may be obtained by applying the Wick's rule to the evaluation of the fermionic 
expectation values, see Appendix \ref{Pfaffians}.

Example. Let us choose $\bt$ as $\bt$ shifted by the variables (\ref{Miwa-choice}) where all $p_i$ are different, 
all $b_i=0$, all $a_i=-1$ and $N_1=N$ even. Then according to relation (\ref{p-bosonization}) and the Wick's rule one can 
write the expectation value as
a Pfaffian of the pairwise expectation values, see Appendix \ref{Pfaffians}.
Then (\ref{det-averages}) is
\be\label{det-averages-Pf}
\int 
\,\prod_{i}^{N}\det\left(1-p_iX \right)^{-1}\, d\mu(X,L,\bt,{\bf s})=
\frac{\prod_{i=1}^N p_i^{(L+1)(2-N)}}{\prod_{i>j}(p_i-p_j)}\Pf \left[ K \right]_{n,m=1,\dots,N}
 \ee
 where the fermionic representation for the tau function $\tau_N(L,{\bf t},{\bf s})$ (\ref{TauN}) yields
 \be
 K_{nm} =\l L|\psi^\dag(p_n)\psi^\dag(p_m)e^{\Phi}|L\r = (p_m-p_n)K_{nm}^*
 \ee
 where $K_{nm}^*$:
 
   \be
 K_{nm}^*= \int_{\mathbb{R}} x^Ly^L
 \frac{|x-y| e^{-\frac 12 x^2-\frac 12 y^2+V(x,\bt)+V(y,\bt)-V(x^{-1},{\bf s})-V(y^{-1},{\bf s})}dx dy }
 {(1-xp_n)(1-yp_n)(1-xp_m)(1-yp_m)}\,\qquad {\rm for \, OE}
  \ee
   \be
 K_{nm}^*= \int_{\mathbb{C}_+} (z-{\bar z})|z|^{2L}
 \frac{\mbox{erfc}\left(\frac{|z-{\bar z}|}{\sqrt 2} \right) 
 e^{-\Re z^2+V(z,\bt)+V({\bar z},\bt)-V(z^{-1},{\bf s})-V({\bar z}^{-1},{\bf s})}d^2z }
 {(1-zp_n)(1-{\bar z}p_n)(1-zp_m)(1-{\bar z}p_m)}
  \ee
   \be
 + \int_{\mathbb{R}} x^Ly^L
 \frac{|x-y| e^{-\frac 12 x^2-\frac 12 y^2+V(x,\bt)+V(y,\bt)-V(x^{-1},{\bf s})-V(y^{-1},{\bf s})}dx dy }
 {(1-xp_n)(1-yp_n)(1-xp_m)(1-yp_m)}\,\quad\qquad \quad {\rm for\, GinOE}
  \ee
      \be
 K_{nm}^*= \int_{\mathbb{R}} x^{2L}
 \frac{ e^{-x^2+2V(x,\bt)-2V(x^{-1},{\bf s})}dx }
 {(1-xp_n)^2(1-xp_m)^2}\,\qquad \qquad \qquad \qquad \qquad \qquad \quad\, {\rm for\, SE}
  \ee
   \be
 K_{nm}^*= \int_{\mathbb{C}_+} 
 |z|^{2L}\frac{(z-{\bar z})^2 e^{-|z|^2+V(z,\bt)+V({\bar z},\bt)-V(z^{-1},{\bf s})-V({\bar z}^{-1},{\bf s})}d^2z }
 {(1-zp_n)(1-{\bar z}p_n)(1-zp_m)(1-{\bar z}p_m)}\,\quad\qquad \quad\, {\rm for \,\, GinSE}
  \ee
Here ${\bar z}$ is the complex conjugated to $z$.
In this example the number of parameters $p_i$ was equal to $N$. In other cases we typically obtain pfaffians of
block matrices. This will be written down in a more detailed text.

\section{Gauss complex Ginibre ensemble and its deformation \label{complex}}

The complex Ginibre ensemble is considered to be easiest and relatively studied and it is quite similar to the well-known 
model of normal matrices \cite{CZ-MWZ}, therefore I will skip details.

{\bf Hermitian random matrices}. First we remind that the ensemble of random Hermitian matrices $X$ with Gauss distribution
for each entry was is known as Gauss unitary ensemble (GUE). The deformation of the measure caused by multiplying 
 the Gauss factor $e^{-\tr X}$ by $e^{\tr V(X,\bt)}$ was related to integrable systems in \cite{GMMMO}. Let us added
 additional deformation parameters $L,{\bf s}$:
  \be
 \int \det (X )^{L} 
e^{\sum_{m=1}\left( (t_m+t'_m)\tr X^m  - (s_m+s'_m)\tr X^{-m}\right)-\tr X^2}d\mu(X)
 \ee
 where $d\mu(X)$ is the Haar measure on the space of Hermitian matrices (see \cite{Mehta} for details). Actually we do not need
 $\bt'$ and ${\bf s}'$ but we keep it to compare the model with the non-Hermitian one.
 The fermionic representation for this model (with ${\bf s}={\bf s}'=0$) was given in \cite{KMMOZ} (see also \cite{ZabrodinGrand}). 
 However to compare the result with the non-Hermitian case it is suitable to use another fermionic construction considered 
 in \cite{Mironych-2-komp} as follows
 \be\label{d-UE}
e_N\l L+N,-N|\,\g({\bf t},{\bf t}')\, 
e^{ \int_{\mathbb{C}}\psi^{(1)}(x)\psi^{\dag(2)}(x)e^{-x^2}dx} 
\,{\bar\g}({\bf s},{\bf s}')\,|
L,0\r
 \ee
with certain independent of deformation parameters factor $e_N$. Here we add the additional deformation  caused by $L,{\bf s}$.
 
{\bf Non-Hermitian random matrices}
Complex restricted Ginibre ensemble with the double deformed measure 
 \be
 \int \det (X)^{L_1}\det(X^\dag )^{-L_2} 
e^{\sum_{m=1}
\left( t_m\tr X^m +t_m'\tr (X^\dag)^m  -s_m\tr X^{-m}-s_m'\tr (X^\dag)^{-m}\right)-\tr XX^\dag}\prod_{i,j}dX_{ij}
 \ee
 (where parameters $L_1,L_2,\bt,\bt',{\bf s},{\bf s}' $ are chosen in a way that the integral is either convergent or 
 may be regularized (see Remark \ref{regul}))
is equal to the following two-component 2D Toda lattice \cite{JM} tau function 
 \be\label{d-GinUE}
f_N\l L_1+N,L_2-N|\,\g({\bf t},{\bf t}')\, 
e^{ \int_{\mathbb{C}}\psi^{(1)}(z)\psi^{\dag(2)}({\bar z})e^{-|z|^2}d^2z} 
\,{\bar\g}({\bf s},{\bf s}')\,|
L_1,L_2\r
 \ee
 The factor $f_N$ is independent of $L_1,L_2,\bt,\bt',{\bf s},{\bf s}'$.

 This type of representation was previously used in \cite{HO2006} in the context of two-matrix and 
 normal matrix models. Perturbation series in deformation parameters for the complex Ginibre ensemble are
 basically the same as found in \cite{HO2006}.

\section*{Acknowledgements}

I am grateful to J. van de Leur, K.Takasaki, T. Shiota, and to J. Harnad
 for helpful discussions of the topic and to E.Kanzieper for an initiating discussion.  I thank
 Kyoto university and CRM in Montreal university for kind hospitality.
 The work was supported by RFBR grants 11-01-00440-а and by Japanese-RFBR grant 10-01-92104 JF.
This work is also partly supported by Grant-in-Aid for Scientific Research
No.~22540186 from the Japan Society for the Promotion
of Science and by the Bilateral Joint Project ``Integrable Systems,
Random Matrices, Algebraic Geometry and Geometric Invariants''
(2010--2011) of the Japan Society for the Promotion of Science and the
Russian Foundation for Basic Research.

\appendix

\section{Appendices}

\subsection{Vertex operators \label{Vertex operators}}

Vertex operators we need are as follows
\be\label{X}
{\hat X}(L,\bt,\lambda):=e^{\sum_{n=1}^\infty \lambda^n t_n}\lambda^{L}
e^{-\sum_{n=1}^\infty \frac{1}{n\lambda^{n}}\frac{\partial}{\partial t_n}}
\,,\quad
{\hat X}^\dag(L,\bt,\lambda):=e^{-\sum_{n=1}^\infty \lambda^n t_n}\lambda^{-L}
e^{\sum_{n=1}^\infty \frac{1}{n\lambda^{n}}\frac{\partial}{\partial t_n}}
 \ee
\be\label{Y}
{\hat Y}(L,{\bf s},\lambda):=e^{-\sum_{n=1}^\infty \lambda^{-n} s_n}\lambda^{L}
e^{\sum_{n=1}^\infty \frac{\lambda^{n}}{n}\frac{\partial}{\partial s_n}}
\,,\quad
{\hat Y}^\dag(L,{\bf s},\lambda):=e^{\sum_{n=1}^\infty \lambda^{-n} s_n}\lambda^{-L}
e^{-\sum_{n=1}^\infty \frac{\lambda^{n}}{n}\frac{\partial}{\partial s_n}}
\ee
Interesting historical fact that the formula which relates fermions to bosons first was found in \cite{PogrebkovSushko}.

The following bosonization relation is useful
 \be\label{p-bosonization}
 \l L+N|\Gamma_+(\bt+\sum_{i=1}^N[p_i])=
 \frac{\l L| \psi^\dag(p_1^{-1})\cdots  \psi^\dag(p_N^{-1})\g_+({\bf t})}{\prod_{i=1}^N p_i^{(L+1)(N-1)} \prod_{i>j}(p_i-p_j)}
 \ee

  Introduce
   \be
   {\hat \Omega}_n(L,\bt)=\res_\lambda \left(\frac{\partial^n {\hat X}(L,\bt,\lambda)}{\partial\lambda^n} 
   {\hat X}^\dag(L,\bt,\lambda)\right)\,,\quad
   {\hat \Omega}_n^*(L,{\bf s})=\res_\lambda  \left({\hat Y}^\dag(L,{\bf s},\lambda) \frac{\partial^n 
   {\hat Y}(L,{\bf s},\lambda)}{\partial\lambda^n}\right)
   \ee

\subsection{Fermions \label{Fermions}}

We shall remind some facts and notations of \cite{JM}. Introduce free fermionic fields
$\psi(z)=\sum_{i\in\mathbb{Z}} \psi_i z^i$, $\psi^\dag(z)=\sum_{i\in\mathbb{Z}} \psi^\dag_{-i-1} z^i$ whose Fourie components 
anti-commute as follows
$\psi_i\psi_j+\psi_j\psi_i=\psi^\dag_i\psi^\dag_j+\psi^\dag_j\psi^\dag_i=0$ and 
$\psi_i\psi^\dag_j+\psi^\dag_j\psi_i=\delta_{i,j}$ where $\delta_{i,j}$ is the Kronecker symbol. We put
 \be
\psi_i|0\r=\psi^\dag_{-i}|0\r =\l 0|\psi_{-i}=\l 0|\psi^\dag_{i}=0
 \ee
where $\l 0|$ and $|0\r$ are left and right vacuum vectors of the fermionic Fock space, $\l 0|\cdot 1 \cdot |0\r=1$. 
Also introduce
 \be
\l n|=
\cases{
\l 0|\psi^\dag_{0}\cdots \psi^\dag_{n-1}\quad {\rm if }\quad n > 0 \cr
\l 0|\psi_{-1}\cdots \psi_{-n}\quad {\rm if }\quad n < 0 
 }
\,,\quad
|n \r =
\cases{
\psi_{n-1}\cdots \psi_{0}|0\r\, \quad {\rm if }\quad n > 0 \cr
\psi^\dag_{-n}\cdots \psi^\dag_{-1}|0\r\, \quad {\rm if }\quad n < 0 
 }
 \ee
Then $\l n|\cdot 1 \cdot |m\r=\delta_{n,m}$.

Following \cite{KvdLbispec} we introduce an additional Fermi mode which we shall denote by 
$\phi$ with properties
\footnote{In notations of \cite{KvdLbispec} our $\psi_n$, $\psi_n^\dag$ and $\phi$ read respectively as
$\psi_{n+\frac 12}$, $\psi^\dag_{n+\frac 12}$ and $\psi_0$} 
 \be
\phi\psi_i+\psi_i\phi=\phi\psi_i^\dag+\psi_i^\dag\phi=0,\quad \phi^2=\frac 12
 \ee
 \be
\phi|0\r=|0\r\frac{1}{\sqrt 2} ,\qquad \l 0|\phi=\frac{1}{\sqrt 2}\l 0|
 \ee
such that $\l L|\phi |L\r=\frac{(-)^L}{\sqrt{2}}$.
 
Now, two-component Fermi fields used in Section \ref{complex} are defined as
  \be
  \psi^{(i)}(z)=\sum_{n\in\mathbb{Z}} z^n\psi_{2n+i}\,,\quad  \psi^{(i)\dag}(z)=
  \sum_{n\in\mathbb{Z}} z^{-n-1}\psi_{2n+1}^\dag
  \ee
Other details about two-component fermions may be found in \cite{JM} or in \cite{HO2006}.
  
We have
 \be
{\hat  X}(L,\bt,\lambda)X^\dag(L,\mu)\l N+L|\Gamma_+(\bt)g\Gamma_-({\bf s})|L\r=
 \l N+L|\Gamma_+(\bt)\psi(\lambda)\psi^\dag(\mu)g\Gamma_-({\bf s})|L\r
 \ee
 \be
{\hat Y}^\dag(L,{\bf s},\mu) Y(\lambda)\l N+L|\Gamma_+(\bt)g\Gamma_-({\bf s})|L\r=
\l N+L|\Gamma_+(\bt)g\psi(\lambda)\psi^\dag(\mu)\Gamma_-({\bf s})|L\r
 \ee

 Then it follows that
  \be
  {\hat \Omega}_n(L,\bt)\l N+L|\Gamma_+(\bt)g\Gamma_-({\bf s})|L\r=
  \l N+L|\Gamma_+(\bt){\tilde \Omega}_n g\Gamma_-({\bf s})|L\r
  \ee
  \be
  {\hat \Omega}_n(L,\bt)\l N+L|\Gamma_+(\bt)g\Gamma_-({\bf s})|L\r=
\l N+L|\Gamma_+(\bt)g \tilde {\Omega}_n\Gamma_-({\bf s})|L\r
  \ee
  where
  \be
  {\tilde \Omega}_n=\res_\lambda \left(\frac{\partial^n \psi(\lambda)}{\partial\lambda^n} \psi^\dag(\lambda)\right)
  \ee
Using the fermionic representation  one may verify that tau functions related to the deformed orthogonal and deformed 
symplectic ensembles obey the constraints
   \be
   \left({\hat \Omega}_n(L,\bt)-{\hat \Omega}^*_n(L,{\bf s})\right)\tau(L,\bt',{\bf s}')=0,\quad n\ge 1 \quad {\rm odd}
   \ee
   where $t'_k=t_k-\frac 12 \delta_{2,k}$ $s'_k=s_k-\frac 12 \delta_{2,k}$ 
   (this shift appears due to the Gauss measure in undeformed ensembles).

    \subsection{The Schur function\label{Schur}}
    
    Consider polynomials $h_n(\bt)$ defined by $e^{\sum_{n=1}^\infty z^nt_n}=\sum_{n=0}^\infty z^nh_n(\bt)$. Then the Schur
    function labeled by a partition $\lambda=(\lambda_1,\dots,\lambda_k>0)$ may be defined as
     $
     s_\lambda(\bt)=\det \left( h_{\lambda_i-i+j}(\bt)\right)_{i,j=1,\dots,k}
     $.

 \subsection{From 2-BKP to TL, BKP, 2-DKP, DKP \label{reductions}}
 
 The general expression for $\Phi$ of (\ref{TauN}) is as follows
 \be\label{Phi-fermion}
 \Phi=\sum_{i,j\in\mathbb{Z}}A_{ij}\psi_i\psi_j +\sum_{i,j\in\mathbb{Z}}B_{ij}\psi^\dag_i\psi_j^\dag 
+ \sum_{i,j\in\mathbb{Z}}D_{ij}\psi_i\psi_j^\dag +\phi\sum_{i\in\mathbb{Z}}a_{i}\psi_i 
+\phi\sum_{i\in\mathbb{Z}}b_{i}\psi^\dag_i
 \ee
 To get TL tau function \cite{JM},\cite{UT} we put all $A_{ij},B_{ij},a_{i},b_{i}$ and $N$ to be zero.
 
 To get BKP \cite{KvdLbispec} we put ${\bf s}=0$.
 
 To get 2-DKP \cite{T-09} we put all $a_i=b_i=0$.
 
 To get DKP \cite{JM} we put ${\bf s}=0$ and all $a_i=b_i=0$.
 
Tau functions (\ref{Theorem1}) and (\ref{Theorem2}) correspond to the case where all $B_{ij},D_{ij},b_i$ vanish (and
for (\ref{Theorem2}) also all $a_i=0$). The case where only $A_{ij}$ (and perhaps $a_i$) are non-vanishing is characterized 
by the condition that the so-called wave functions
  \be
  w^{(\infty)}(N,L,\bt,{\bf s},\lambda)=\frac{{\hat X}(L,\bt,\lambda) \tau_N(L,\bt,{\bf s})}{\tau_N(L,\bt,{\bf s})}=
  \lambda^L e^{V(\lambda,\bt)-V(\lambda^{-1},{\bf s})}P_N(L,\bt,{\bf s},{\bf},\lambda)
  \ee
  \be
  w^{(0)}(N,L,\bt,{\bf s},\lambda)=\frac{{\hat Y}^\dag(L,{\bf s},\lambda) \tau_N(L,\bt,{\bf s})}{\tau_N(L,\bt,{\bf s})}=
  \lambda^Le^{V(\lambda,\bt)-V(\lambda^{-1},{\bf s})} Q_N(L,\bt,{\bf s},\lambda)
  \ee
  are equal, and $P_N=Q_N$ are polynomials in $\lambda$ of the order $N$.

 \subsection{Pfaffians\label{Pfaffians}}

If $A$ an anti-symmetric matrix of an odd order its determinant
vanishes. For even order, say $k$, the following multilinear form
in $A_{ij},i<j\le k$
 \be\label{Pf''}
\Pf [A] :=\sum_\sigma
{\sgn(\sigma)}\,A_{\sigma(1),\sigma(2)}A_{\sigma(3),\sigma(4)}\cdots
A_{\sigma(k-1),\sigma(k)}
 \ee
where sum runs over all permutation restricted by
 \be
\sigma:\,\sigma(2i-1)<\sigma(2i),\quad\sigma(1)<\sigma(3)<\cdots<\sigma(k-1),
 \ee
 coincides with the square root of $\det A$ and is called the
 {\em Pfaffian} of $A$, see, for instance \cite{Mehta}. As one can see the Pfaffian  contains
 $1\cdot  3\cdot 5\cdot \cdots \cdot(k-1)=:(k-1)!!$ terms.
 
 \paragraph{ Wick's relations.} Let each of $w_i$ be a linear
combination of Fermi operators:
  \[
{\hat w}_i=\sum_{m\in\mathbb{Z}}\,v_{im}\psi_m\,+\,
\sum_{m\in\mathbb{Z}}\,u_{im}\psi^\dag_m\,
 ,\quad i=1,\dots,n
 \]
  Then the Wick formula is
  \be\label{Wick} \l l|{\hat w}_1\cdots {\hat w}_n |l\r =
\cases{
\Pf\left[ A \right]_{i,j=1,\dots,n} \quad {\rm if\,\,n\,is\,even}  \cr
0 \qquad\qquad\qquad\quad\, \mbox{otherwise}
 }
  \ee  
    where $A$ is $n$ by $n$ antisymmetric matrix with entries
 $ A_{ij}\, = \,\l l|{\hat w}_i {\hat w}_j|l\r\, ,\quad i<j$.

 \subsection{Hirota equations \label{Hirota}}
 Hirota equations for the large BKP hierarchy were written in \cite{KvdLbispec}. For 2-BKP hierarchy 
 Hirota equations are as follows \cite{OST-I}
 \bea\label{Hirota-2lBKPtau}
  \oint\frac{dz}{2\pi i}z^{N'+l'-N-l-2}e^{V(\bt'-\bt,z)}
  \tau_{N'-1}(l',\bt'-[z^{-1}],{\bf s}')
  \tau_{N+1}(l,\bt+[z^{-1}],{\bf s}) \nonumber\\
+ \oint\frac{dz}{2\pi i}z^{N+l-N'-l'-2}e^{V(\bt-\bt',z)}
  \tau_{N'+1}(l',\bt'+[z^{-1}],{\bf s}')
  \tau_{N-1}(l,\bt-[z^{-1}],{\bf s}) \nonumber\\
= \oint\frac{dz}{2\pi i}z^{l'-l}e^{V({\bf s}'-{\bf s},z^{-1})} 
  \tau_{N'-1}(l'+1,\bt',{\bf s}'-[z])
  \tau_{N+1}(l-1,\bt,{\bf s}-[z]) \nonumber \\
+ \int\frac{dz}{2\pi i}z^{l-l'}e^{V({\bf s}'-{\bf s},z^{-1})}
  \tau_{N'+1}(l'-1,\bt',{\bf s}'+[z])
  \tau_{N-1}(l+1,\bt,{\bf s}+[z]) \nonumber\\
+ \frac{(-1)^{l'+l}}{2}(1-(-1)^{N'+N})
  \tau_{N'}(l',\bt',{\bf s}')\tau_N(l,\bt,{\bf s}) 
\eea
 The difference Hirota equation may be obtained from the previous one \cite{OST-I}
 \bea
  - \frac{\beta}{\alpha-\beta}
    \tau_N(l,\bt+[\beta^{-1}])\tau_{N+1}(l,\bt+[\alpha^{-1}]) 
  - \frac{\alpha}{\beta-\alpha}
    \tau_N(l,\bt+[\alpha^{-1}])\tau_{N+1}(l,\bt+[\beta^{-1}]) 
  \nonumber\\
  + \frac{1}{\alpha\beta}
    \tau_{N+2}(l,\bt+[\alpha^{-1}]+[\beta^{-1}])\tau_{N-1}(l,\bt) 
  = \tau_{N+1}(l,\bt+[\alpha^{-1}]+[\beta^{-1}])\tau_N(l,\bt). 
\eea

\subsection{Integrals over orthogonal and symplectic groups \label{integrals-over} }

Using
 \be\label{} \int_{O\in \mathbb{O}(N)} s_\lambda(O) d_*O =
\cases{
1 \quad \lambda\,\mbox{is\,even}  \cr
0 \quad\, \mbox{otherwise}
 }\,,\qquad
\int_{S\in \mathbb{S}p(N)} s_\lambda(S) d_*S =
\cases{
1 \quad \lambda^{tr}\,\mbox{is\,even}  \cr
0 \quad\, \mbox{otherwise}
 }
  \ee  
we get
 \be
J_1(\bt,N):=\int_{O\in \mathbb{O}(N)} e^{\sum_{m=1}^\infty t_m\Tr O^m} d_*O = 
\sum_{\lambda\,{\rm even}\atop \ell(\lambda)\le N} s_\lambda(\bt)
 \ee
 \be
J_2(\bt,N):=\int_{S\in \mathbb{S}p(2n)} e^{\sum_{m=1}^\infty t_m\Tr S^m} d_*S = 
\sum_{\lambda^{tr}\,{\rm even}  \atop \ell(\lambda)\le 2n} s_\lambda(\bt)
 \ee

The right hand sides were obtained in \cite{OST-I} as examples of the BKP tau function. Thus integrals $ J_1(\bt,N)$ and
$J_2(\bt,N)$ are tau functions, $N$ and $\bt$ being BKP higher times.

\end{document}